\documentclass[aps,prc,twocolumn,superscriptaddress,showpacs,floatfix,nofootinbib]{revtex4-1}
\usepackage{amsmath,graphicx}
\usepackage{color}

\begin{document}

\title{Eccentricity distributions in nucleus-nucleus collisions}

\author{Li Yan}
\author{Jean-Yves Ollitrault}
\affiliation{
CNRS, URA2306, IPhT, Institut de physique th\'eorique de Saclay, F-91191
Gif-sur-Yvette, France} 
\author{Arthur M. Poskanzer}
\affiliation{Lawrence Berkeley National Laboratory, Berkeley,
California, 94720}
\date{\today}

\begin{abstract}
We propose a new parametrization of the distribution of the initial
eccentricity 
in a nucleus-nucleus collision at a fixed centrality, which we name
the Elliptic Power distribution. 
It is a two-parameter distribution, where one of the parameters
corresponds to the 
intrinsic eccentricity, while the other parameter controls the
magnitude of eccentricity fluctuations.  
Unlike the previously used Bessel-Gaussian distribution, which 
becomes worse for more peripheral collisions, 
the new Elliptic Power distribution fits several Monte Carlo models
of the initial state for all centralities. 
\end{abstract}


\maketitle
\section{Introduction}
Elliptic flow, $v_2$, is a crucial observable of heavy-ion collisions: 
the large magnitude of $v_2$ at RHIC~\cite{Ackermann:2000tr,Adler:2003kt}
and LHC~\cite{Aamodt:2010pa,ATLAS:2011ah,Chatrchyan:2012ta} 
provides the strongest evidence that a low-viscosity fluid is formed in these 
collisions~\cite{Romatschke:2007mq,Luzum:2008cw}. 
Elliptic flow  is determined to a good approximation by linear
response to the initial eccentricity $\varepsilon_2$, which 
quantifies the spatial azimuthal anisotropy of the fireball created
right after the collision~\cite{Alver:2006wh}.
This initial eccentricity comes from two effects: 
first, the overlap area between the colliding nuclei has the shape of an almond in non-central 
collisions, where the smaller dimension of the almond is parallel to the
reaction plane.  
This results in an eccentricity which becomes
larger as impact parameter increases,  
and whose magnitude is model dependent~\cite{Hirano:2005xf,Lappi:2006xc}.
Second, even in central collisions, there is a sizable eccentricity due to 
quantum fluctuations in wave functions of incoming 
nuclei~\cite{Miller:2003kd,Alver:2006wh}, and to the probabilistic
nature of energy deposition in nucleon-nucleon collisions. 
The magnitude of these eccentricity fluctuations  is again a
model-dependent issue, which involves the dynamics of the collision at early 
times~\cite{Miller:2007ri,Flensburg:2011wx,Dumitru:2012yr,Schenke:2012hg}. 
The goal of this paper is to show that the distribution of the initial
eccentricity is to some extent independent of model details. 
More precisely, it can be written to a good  approximation as a 
universal function of two parameters, where one of the parameters
corresponds to the reaction plane eccentricity, and the other
parameter characterizes the magnitude of fluctuations. 
Information on the initial state is thus encoded in two numbers.

The initial eccentricity $\varepsilon_2$ 
is defined in every event from the initial energy   
density profile (see below Sec.~\ref{sec:ellipticpower})
and thus carries information about how energy is deposited in the early 
stages of the collision. 
There are several models of the initial density profile and its fluctuations, which are typically 
implemented through Monte Carlo simulations. 
The Monte Carlo Glauber 
model is used  in many event-by-event hydrodynamic 
calculations~\cite{Holopainen:2010gz,Schenke:2010rr,Qiu:2011iv,Bozek:2011if}: 
in this model, one assumes that the energy is localized around 
each wounded nucleon. Other Monte Carlo models of the initial state are inspired by saturation 
physics~\cite{Drescher:2007ax,Flensburg:2011wx,Dumitru:2012yr,Schenke:2012hg} 
and have also been used as initial conditions in hydrodynamic calculations~\cite{Schenke:2012wb}. 
Another approach is to use an event generator from particle
physics~\cite{Andrade:2006yh} or 
a transport calculation~\cite{Lin:2004en,Petersen:2008dd} to model the 
initial dynamics.  
Each  Monte Carlo model returns a probability distribution for $\varepsilon_2$ at a given centrality.

A simple parametrization of the distribution of $\varepsilon_2$, 
usually referred to as 
the Bessel-Gaussian distribution~\cite{Voloshin:2008dg}, 
was proposed in~\cite{Voloshin:2007pc}. 
It works well for nucleus-nucleus collisions at moderate impact parameters, 
but fails for more peripheral collisions and/or small systems such as proton-nucleus collisions. 
The reason why it fails can be traced back to the fact that it does
not take into account the fact that, by definition, $\varepsilon_2<1$ in
every event.  
A new Power distribution was recently introduced~\cite{Yan:2013laa}
which well describes eccentricity distributions when there are only
flow fluctuations 
(see also \cite{Bzdak:2013rya,Bzdak:2013raa}), and satisfies $\varepsilon_2<1$ by construction.
In  Sec.~\ref{sec:ellipticpower}, we propose a generalization of this
result:  we take into account the  eccentricity 
in the reaction plane by distorting the Power distribution into an
Elliptic Power distribution. 
This new, two-parameter distribution reduces to the 
Power distribution for an azimuthally-symmetric system. 
In Sec.~\ref{sec:mcmodels}, we use the Elliptic Power distribution to fit 
the distribution of $\varepsilon_2$ in Pb+Pb collisions calculated by Monte Carlo methods for several models and for 
all centralities. 
We also show that the Elliptic Power distribution reproduces the magnitude of eccentricity fluctuations, and the 
cumulants of the distribution of $\varepsilon_2$. 

\section{The Elliptic Power distribution}
\label{sec:ellipticpower}

\begin{figure*}
\includegraphics[width=.8\linewidth]{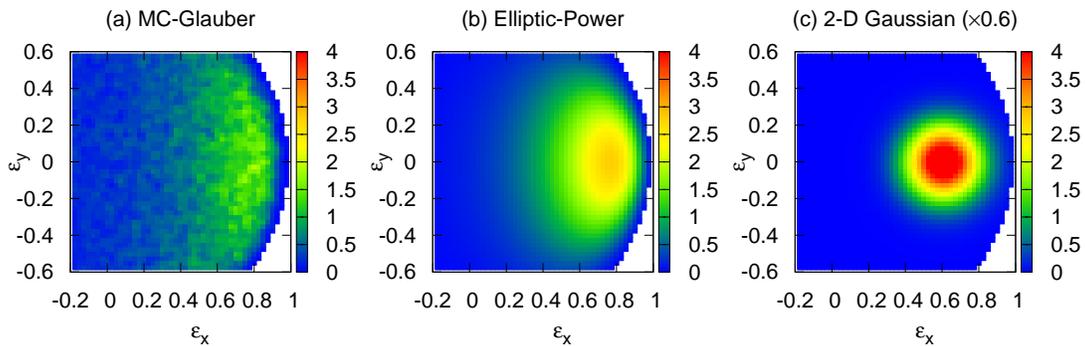}
\caption{ (Color online) 
(a) Two-dimensional plot of the distribution of
$(\varepsilon_x,\varepsilon_y)$, with $n=2$, in a Monte Carlo
Glauber~\cite{Alver:2008aq} simulation of Pb+Pb collisions at 2.76~TeV
per nucleon pair, in the 75-80\% centrality range. 
40000 events are generated in this centrality. 
(b) Fit using the Elliptic Power distribution 
Eq.~(\ref{ellipticpower2d}) with $\varepsilon_0=0.61$ and $\alpha=3.3$. 
(c) Fit using an isotropic two-dimensional
Gaussian~\cite{Voloshin:2007pc},  
corresponding to Eq.~(\ref{BG2d}) 
with $\varepsilon_0=0.61$ and $\sigma_x=\sigma_y=0.10$. The z-axis of the right panel has been reduced by 0.6 to match the others.
All fit parameters are obtained by standard $\chi^2$ fits 
to the distribution of $\varepsilon_2$, see Fig.~\ref{1d} (a).
}
\label{2d}
\end{figure*}

\subsection{Definition and example} 
\label{sec:example}

The initial anisotropy in harmonic $n$ 
is defined in every event 
by~\cite{Teaney:2010vd} 
\begin{equation}
\label{defepsilon}
\varepsilon_n e^{in\psi_n}
\equiv -\frac{\int r^n e^{ni\varphi}\rho(r,\varphi)r{\rm d}r{\rm d}\varphi}
{\int r^n\rho(r,\varphi)r{\rm d}r{\rm d}\varphi}.
\end{equation}
where $\rho(r,\varphi)$ is the energy density near midrapidity shortly
after the collision, and  
$(r,\varphi)$ are polar coordinates in the transverse plane, in a
centered coordinate system, where $\varphi=0$ is the orientation of the
reaction plane. 
In most of this paper, we focus on the second harmonic $n=2$. 
$\varepsilon_2$ is often referred to as the ``participant
eccentricity'' and $\psi_2$ as the ``participant plane''. 
This terminology refers to Monte Carlo Glauber
models~\cite{Miller:2007ri}, in the context of which these concepts
were first introduced~\cite{Alver:2006wh}.
Note that $0\le\varepsilon_n\le 1$ by definition. 

The initial anisotropy can also be written in cartesian coordinates:
\begin{equation}
\label{cartesian}
\varepsilon_n e^{in\psi_n}=\varepsilon_x+i\varepsilon_y.
\end{equation}
$\varepsilon_x$ is the anisotropy in the reaction plane. 
For a symmetric density profile satisfying 
$\rho(r,\varphi)=\rho(r,-\varphi)$, the definition
Eq.~(\ref{defepsilon}) implies $\varepsilon_y=0$, which in turn implies
$\psi_n=0$: the participant plane coincides with the reaction
plane. This is no longer true in the presence of fluctuations. 

Figure~\ref{2d} (a) displays the distribution of
$(\varepsilon_x,\varepsilon_y)$, for $n=2$, obtained in a Monte Carlo Glauber
simulation of  Pb+Pb collisions in the 75-80\% centrality range. 
In this simulation, centrality is defined according the to number of
participants.
The maximum of the distribution is at a positive value of 
$\varepsilon_x$, reflecting the large reaction plane eccentricity. 
Fluctuations around the most probable value are large. 
They display characteristic features:
\begin{enumerate}
\item The width around the maximum is larger along the $y$ axis than along
the $x$ axis. 
\item The distribution of $\varepsilon_x$ is left skewed 
with a steeper decrease to the right of the maximum toward a cut-off at $\varepsilon_2 = 1$. 
\end{enumerate}
The usual Bessel-Gaussian parametrization~\cite{Voloshin:2007pc}  
assumes that fluctuations are Gaussian and isotropic
and therefore misses both features (see Fig.~\ref{2d} (c)). 
These features can be traced back to the constraint that the
support of the distribution is the unit disk $\varepsilon_2\le 1$. Our
goal in this paper is to derive a generic distribution with these 
features. 

\subsection{Two-dimensional distribution} 
\label{sec:ep2d}

In Ref.~\cite{Ollitrault:1992bk},\footnote{See Eq.~(3.9)
  of~\cite{Ollitrault:1992bk}. The 
  result was derived for the eccentricity in momentum space, but the
  algebra is identical.}
  an exact expression for the
distribution of $(\varepsilon_x,\varepsilon_y)$ for $n=2$ was derived under the
following assumptions:
\begin{itemize}
\item The energy profile is a superposition of 
$N$ pointlike, identical sources: $\rho({\bf
  x})\propto\sum_{j=1}^N\delta({\bf x}-{\bf x}_j)$, where ${\bf x}_j$
  denotes the transverse position of the sources.
\item The positions of the sources ${\bf x}_j$ are independent. 
\item The distribution of ${\bf x}_j$ is a 2-dimensional Gaussian,
  where the widths along $x$ and $y$ may differ. Here we denote by 
$\varepsilon_0\equiv\langle y_j^2-x_j^2\rangle/\langle
  y_j^2+x_j^2\rangle$ the {\it ellipticity\/} parameter, corresponding
  to the eccentricity of the distribution of sources in the reaction
plane. It satisfies $|\varepsilon_0|\le 1$.  
\end{itemize}
Under these conditions, the distribution of
$(\varepsilon_x,\varepsilon_y)$ is  
\begin{equation}
p(\varepsilon_x,\varepsilon_y)=
\frac{\alpha}{\pi}
  (1-\varepsilon_0^2)^{\alpha+\frac{1}{2}}\frac{(1-\varepsilon_x^2-\varepsilon_y^2)^{\alpha-1}}
{(1-\varepsilon_0\varepsilon_x)^{2\alpha+1}},
\label{ellipticpower2d}
\end{equation}
where $\alpha=(N-1)/2$. 
This probability distribution is 
normalized: 
$\int p(\varepsilon_x,\varepsilon_y)d\varepsilon_xd\varepsilon_y=1$, where
integration runs over the unit disk $\varepsilon_x^2+\varepsilon_y^2\le 1$. 

In this paper, we argue that Eq.~(\ref{ellipticpower2d}), which we
name the Elliptic Power distribution, provides a good
fit to {\it all\/} models of the initial state. 
This success can be ascribed to the fact that 
the natural support of the Elliptic Power distribution is the unit
disk: this is a major advantage over previous parametrizations. 
We treat both the {\it ellipticity\/} $\varepsilon_0$ and the 
{\it power\/} $\alpha$ as fit parameters.  
In particular, we allow for arbitrary real, positive values of 
$\alpha$ (as opposed to integer or half-integer). 

For $\varepsilon_0=0$, the distribution Eq.~(\ref{ellipticpower2d}) is
azimuthally symmetric:  
\begin{equation}
p(\varepsilon_x,\varepsilon_y)=
\frac{\alpha}{\pi}
(1-\varepsilon_x^2-\varepsilon_y^2)^{\alpha-1}.
\label{power2d}
\end{equation}
This is the one-parameter Power distribution introduced in Ref.~\cite{Yan:2013laa},
which was shown to fit Monte Carlo results when the
eccentricity is solely created by fluctuations, as for instance in
p-p collisions\footnote{
In p-p collisions, as indicated in \cite{Yan:2013laa} via comparisons to DIPSY model, 
fluctuation-induced eccentricity~\cite{CasalderreySolana:2009uk} plays a 
dominant role irrespective of the effect of non-zero
impact parameter~\cite{Prasad:2009bx}. 
}
or p-Pb collisions. The power parameter $\alpha$ 
quantifies the magnitude of fluctuations: the
smaller $\alpha$, the larger the fluctuations. 

When the ellipticity $\varepsilon_0$ is
positive, the
denominator of Eq.~(\ref{ellipticpower2d}) breaks azimuthal symmetry and favors larger values of $\varepsilon_x$. 
The mean eccentricity in the reaction plane
$\varepsilon_{\rm{RP}}\equiv\langle\varepsilon_x\rangle$ is derived in
Appendix~\ref{sec:appendix} as a function of $\varepsilon_0$ and
$\alpha$. 
Because of fluctuations, it is not strictly equal to the eccentricity of
the underlying distribution, $\varepsilon_0$~\cite{Bhalerao:2006tp}. 
It is in general smaller, and coincides with $\varepsilon_0$ only in
the limit $\alpha\gg 1$. 

A fit to Monte Carlo Glauber results using the Elliptic Power
distribution is displayed in Fig.~\ref{2d} (b). 
The fit is not perfect. Specifically, the maximum density is 
slightly overestimated, while the  
width of the $\varepsilon_y$ distribution is slightly underestimated. 
Note that there are several differences between the ideal case 
considered in~\cite{Ollitrault:1992bk} and the actual Glauber
calculation, specifically:  the correlations between the participants, 
the fact that their distribution in the transverse plane is not a Gaussian, 
and the recentering correction. 
We have checked that switching off the recentering correction does not
make agreement significantly better. 
Despite these imperfections, the Elliptic Power distribution captures
both features pointed out at the end of Sec.~\ref{sec:example},
namely, a larger width along the $y$ axis, and a steeper decrease to the
right of the maximum. 

The Elliptic Power distribution can be somewhat simplified 
in the limit $\alpha\gg 1$, corresponding to a large system with small fluctuations. 
To leading order in $1/\alpha$, 
Eq.~(\ref{ellipticpower2d}) reduces to a two-dimensional elliptic Gaussian
distribution:
\begin{equation}
p(\varepsilon_x,\varepsilon_y)=\frac{1}{2\pi\sigma_x\sigma_y}
\exp\left(-\frac{(\varepsilon_x-\varepsilon_0)^2}{2\sigma_x^2}
-\frac{\varepsilon_y^2}{2\sigma_y^2}\right).
\label{BG2d}
\end{equation}
The maximum lies on the $x$-axis at $\varepsilon_x=\varepsilon_0$ and
the widths are given by 
\begin{eqnarray}
\label{sigmaxy}
\sigma_x&=&\frac{1-\varepsilon_0^2}{\sqrt{2\alpha}}\cr
\sigma_y&=&\sqrt{\frac{1-\varepsilon_0^2}{2\alpha}}.
\end{eqnarray}
In general, the Gaussian is more elongated along the $y$ axis, that
is, $\sigma_x<\sigma_y$, which corresponds
to the first of the two properties listed in Sec.~\ref{sec:example}. 
It is symmetric around its maximum and therefore does not possess the
second property, namely, the skewness along the $x$ axis. 
This property only appears as a next-to-leading correction 
of order $1/\alpha$, which is derived in 
Appendix~\ref{sec:asymptotic}.

The usual isotropic Gaussian distribution introduced in
Ref.~\cite{Voloshin:2007pc} is obtained by 
setting $\sigma_x=\sigma_y=\sigma$ in 
Eq.~(\ref{BG2d}). 
This parametrization misses both properties and is therefore less
accurate than our new Elliptic Power distribution, 
as can be seen in Fig.~\ref{2d} (c). 
In particular, it overestimates the density at the maximum by a factor
larger than  2. 

\subsection{Radial distribution}
\begin{figure}
\includegraphics[width=.7\linewidth]{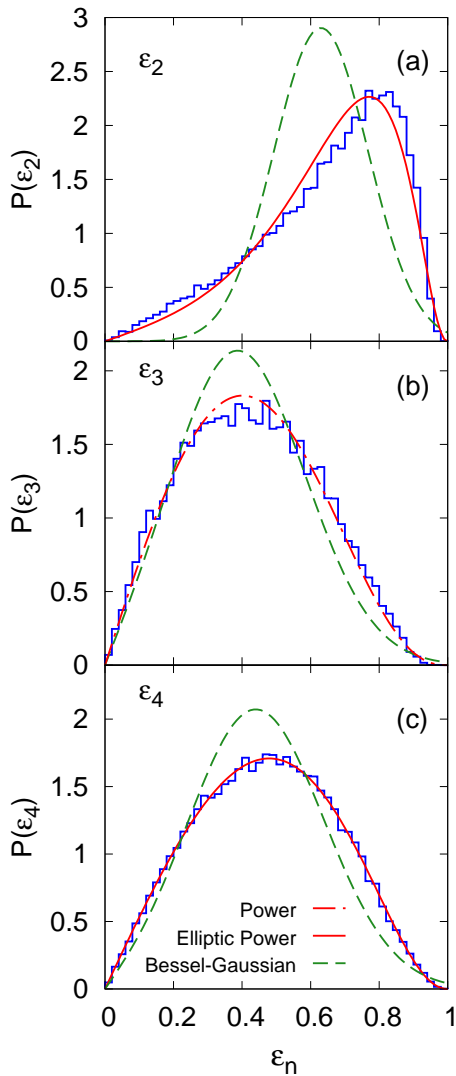}
\caption{ (Color online) 
Distribution of $\varepsilon_n$ in 
75--80\% central Pb-Pb collisions. 
(a): $\varepsilon_2$, (b): $\varepsilon_3$, (c): $\varepsilon_4$.
Histograms are Monte Carlo Glauber simulations  (same as in
Fig.~\ref{2d} (a)). 
Dashed lines are Bessel-Gaussian fits using Eq. ~(\ref{BG}). 
Full lines are Elliptic Power fits using Eq.~(\ref{ellipticpower}) for 
$\varepsilon_2$ (parameters as in Fig.~\ref{2d} (b)) and $\varepsilon_4$ 
($\alpha=3.2$ and $\varepsilon_0=0.22$).
The dash-dotted line for $\varepsilon_3$ is 
a Power fit using Eq.~(\ref{power}) ($\alpha=3.6$). }
\label{1d}
\end{figure}

Since the orientation of the reaction plane is not directly accessible
experimentally, the magnitude of the eccentricity $\varepsilon_n$
matters more than its phase $\varphi\equiv n\psi_n$. 
Monte Carlo simulations of the initial state typically return a
probability distribution $P(\varepsilon_n)$ for each
centrality  \cite{Aad:2013xma,Schenke:2013aza,Yan:2013laa}.
It is obtained by transforming $p(\varepsilon_x,\varepsilon_y)$ 
to polar coordinates and integrating over the azimuthal angle:
\begin{equation}
\label{cartesiantopolar}
P(\varepsilon_n)\equiv \varepsilon_n\int_0^{2\pi}
p(\varepsilon_n\cos\varphi,\varepsilon_n\sin\varphi)d\varphi.
\end{equation}
It is normalized by construction:
$\int_0^1P(\varepsilon_n)d\varepsilon_n=1$. 
Inserting Eq.~(\ref{ellipticpower2d}) into
Eq.~(\ref{cartesiantopolar}) and using the symmetry of the integrand
under $\varphi\to-\varphi$, one obtains 
\begin{eqnarray}
P(\varepsilon_n)&=&2\alpha\varepsilon_n(1-\varepsilon_n^2)^{\alpha-1}(1-\varepsilon_0^2)^{\alpha+\frac{1}{2}}
\cr
&&\times
\frac{1}{\pi}\int_0^{\pi} (1-\varepsilon_0\varepsilon_n\cos\varphi)^{-2\alpha-1}d\varphi . 
\label{ellipticpower}
\end{eqnarray}
The integral can be carried out analytically to give
\begin{eqnarray}
P(\varepsilon_n)&=&2\varepsilon_n\alpha (1-\varepsilon_n^2)^{\alpha-1} 
(1-\varepsilon_n\varepsilon_0)^{-1-2 \alpha} 
(1-\varepsilon_0^2)^{\alpha+\frac{1}{2}}\times \cr
&&
{_2}F_1\left(\frac{1}{2},1+2\alpha;1;\frac{2\varepsilon_n\varepsilon_0}{\varepsilon_n\varepsilon_0-1}\right).
\end{eqnarray}
However, if the hypergeometric function is not available, or not defined everywhere needed, the integral over angles in Eq.~(\ref{ellipticpower})
may be carried out numerically.\footnote{A fast and accurate method is to evaluate the Riemann sum over 
$n$ equally spaced angles $\varphi_k=(2k-1)\pi/(2 n)$, where $k=1,\cdots, n$.  
Excellent accuracy is obtained with $n=50$ integration points.}

For $\varepsilon_0=0$, Eq.~(\ref{ellipticpower}) reduces to 
\begin{equation}
P(\varepsilon_n)=2\alpha\varepsilon_n(1-\varepsilon_n^2)^{\alpha-1},
\label{power}
\end{equation}
which is the ``Power'' distribution~\cite{Yan:2013laa}. In the limit $\alpha\gg 1$ this becomes a Gaussian.
In the limit $\alpha\gg 1$ and $\varepsilon_0\ll 1$, 
Eqs.~(\ref{BG2d}) and (\ref{cartesiantopolar}) give
\begin{equation}
P(\varepsilon_n)=\frac{\varepsilon_n}{\sigma^2}\exp\left(-\frac{\varepsilon_n^2+\varepsilon_0^2}{2\sigma^2}\right)
I_0\left(\frac{\varepsilon_0\varepsilon_n}{\sigma^2}\right),
\label{BG}
\end{equation}
which is the usual Bessel-Gaussian
distribution~\cite{Voloshin:2007pc}, where we have defined
$\sigma\equiv 1/\sqrt{2\alpha}$. 
Note that if $\sigma_x\not=\sigma_y$, the two-dimensional elliptic Gaussian
distribution Eq.~(\ref{BG2d}) does not give the Bessel-Gaussian
distribution upon integration over $\varphi$~\cite{Ollitrault:1992bk}.

Figure~\ref{1d} (a) presents the histogram of $\varepsilon_2$ 
obtained by integrating the results of Fig.~\ref{2d} (a) over azimuthal
angle. 
The fit using the Elliptic Power distribution is clearly much better
than the fit using the Bessel-Gaussian.\footnote{Note that the Bessel-Gaussian fit is very sensitive to the weights used in
the fitting procedure.   
Our standard $\chi^2$ fit gives a large weight to the last bin because the 
Bessel-Gaussian does not go to zero at $\varepsilon_2=1$.
The Elliptic Power distribution gives a much better fit than the
Bessel-Gaussian, irrespective of the details of the fit procedure.}

For sake of completeness, Fig.~\ref{1d} (b) and (c) also display
the distributions of higher order Fourier harmonics $\varepsilon_3$ and 
$\varepsilon_4$. 
The initial {\it triangularity\/}  $\varepsilon_3$ 
acts as a seed for triangular anisotropy~\cite{Alver:2010gr}, in the same way as 
the initial eccentricity is the origin of elliptic anisotropy. 
Since triangularity is solely created by fluctuations, the
distribution of $\varepsilon_3$ is well reproduced by the 
single-parameter Power distribution, Eq.~(\ref{power})~\cite{Yan:2013laa}.
If the two-parameter Elliptic Power distribution is used for $\varepsilon_3$, the $\varepsilon_0$ parameter comes out to be essentially zero.
The one-parameter fit is significantly better than the 
two-parameter Bessel-Gaussian fit, Eq.~(\ref{BG}). 
Note that the values of $\alpha$ are not necessarily the same for 
ellipticity and triangularity.
The distribution of the fourth harmonic $\varepsilon_4$ is 
well fitted by the Elliptic Power distribution. The resulting
value of $\varepsilon_0$ is significantly smaller than for the distribution
of $\varepsilon_2$. 
Note, however, that the $\varepsilon_0$ for $\varepsilon_4$ is not the only origin of anisotropy 
in the corresponding harmonic, due to large nonlinear terms in the 
hydrodynamic response~\cite{Teaney:2010vd,Gardim:2011xv,Teaney:2012ke}.

\begin{figure*}
\includegraphics[width=0.8\linewidth]{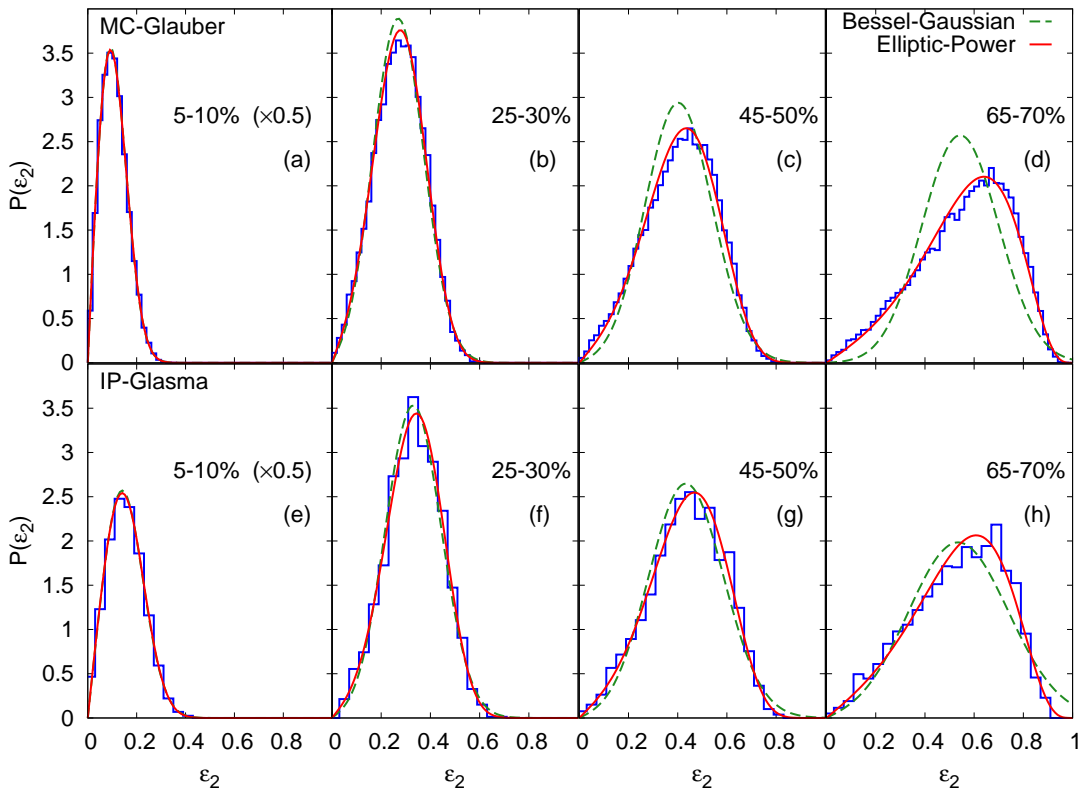}
\caption{ (Color online) 
Histograms of the distribution of 
$\varepsilon_2$ in Pb+Pb collisions at 2.76~TeV using the 
PHOBOS Monte Carlo Glauber~\cite{Alver:2008aq}  (panels (a) to (d)) and the
IP-Glasma~\cite{Schenke:2013aza,Schenke:2014tga} (panels (e) to (h)) for four centrality bins 
(with decreasing centrality or increasing centrality percentile from left to right). 
Solid curves are fits using the Elliptic Power distribution, 
Eq.~(\ref{ellipticpower}), dashed curves are fits using the Bessel-Gaussian
distribution, Eq.~(\ref{BG}). 
Each bin contains $\sim$40000 events for the Glauber simulation,
and $\sim$2000 events for the IP-Glasma, which explains the larger 
statistical fluctuations for the bottom row even though the bins are twice as wide. 
The area under the curves is $0.5$ for the 5-10\% centrality bin and $1$ for the other bins. 
}  
\label{histo}
\end{figure*}

\section{Analyzing Monte Carlo models of Pb+Pb collisions}
\label{sec:mcmodels}

\begin{figure}
\includegraphics[width=\linewidth]{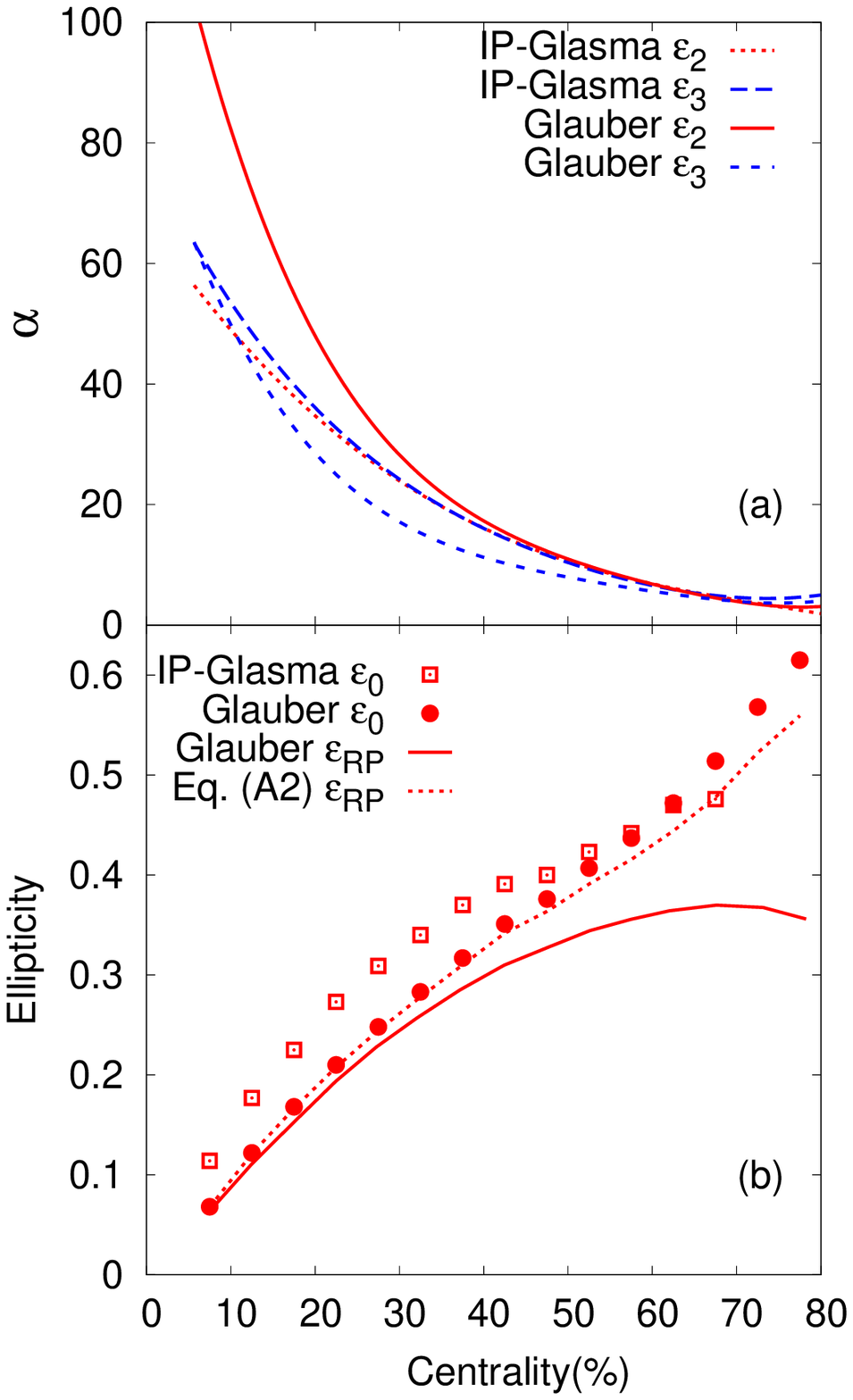}
\caption{ (Color online)
(a):
Power parameter $\alpha$ from Elliptic Power fits to the eccentricity
distribution  
and Power fits to the triangularity distribution  [cf. Fig.~\ref{1d} (b)], 
as a function of centrality for the 
Monte Carlo Glauber
model~\cite{Alver:2008aq} 
and the IP-Glasma model~\cite{Schenke:2013aza}. 
(b): 
Symbols correspond to the value of the 
ellipticity parameter $\varepsilon_0$. 
Lines correspond to the mean eccentricity in the reaction plane 
$\varepsilon_{\rm{RP}}$ for the Glauber model, 
either calculated directly (full line),
or estimated using the Elliptic Power distribution Eq.~(\ref{epsRP}) (dotted line).
}  
\label{parameters}
\end{figure}

\subsection{Histograms}

We now argue that the Elliptic Power distribution always gives good
fits to distributions of $\varepsilon_2$ in nucleus-nucleus collisions. 
Figure~\ref{histo} presents the histogram of $\varepsilon_2$ in Pb+Pb at
2.76~TeV in several centrality bins, obtained using the Monte Carlo
Glauber (panels (a) to (d))~\cite{Alver:2008aq} 
and the IP-Glasma (panels (e) to (h))~\cite{Schenke:2012hg} 
models,\footnote{The centrality is defined according to the number of participants
in the Glauber model and according to the gluon multiplicity~\cite{Gale:2012rq} in the IP-Glasma model.}
together with fits using the Elliptic Power and the
Bessel-Gaussian distributions. 
Both distributions are able to fit both
models in the 5-10\% centrality bin. Bessel-Gaussian fits become 
worse as the centrality percentile is increased, while Elliptic
Power fits are excellent for both models and for all centralities. 
Only four centrality bins are shown in Fig.~\ref{histo} for sake of
illustration, but we have checked that the fits are as good for the other
centralities. 
For the most central bin (0-5\%), however, the fit parameters are
strongly correlated and cannot be determined independently. 
This can be understood as follows: for central collisions, the Elliptic
Power distribution is very close to a Bessel-Gaussian distribution,  
Eq.~(\ref{BG}). 
Now, to order $\varepsilon_0^2$, this distribution is invariant under the
transformation
$(\sigma^2,\varepsilon_0)\to(\sigma^2+\varepsilon_0^2/2,0)$, i.e, the 
dependence on $\varepsilon_0$ can be absorbed into a redefinition of
the width $\sigma$. Therefore one cannot fit $\varepsilon_0$ and
$\sigma$ independently when $\varepsilon_0$ is too small and one can actually use the one-parameter Power distribution. 

The two models plotted in Fig.~\ref{histo} represent two extremes in
the landscape of initial-state models. The PHOBOS Monte Carlo model is
the simplest model including fluctuations: all participant nucleons
are treated as identical, pointlike sources of energy. 
By contrast, in the IP-Glasma model, the energy density is treated as a
continuous field and contains nontrivial fluctuations at the
subnucleonic level. 
The Elliptic Power distribution is able to fit both extremes. 
We have explicitly checked that it works well also for the MC-KLN
model~\cite{Drescher:2007ax}.
We therefore conjecture that it provides a good fit to all 
Monte Carlo models of initial conditions.

\subsection{Power parameter and Ellipticity}

The Elliptic Power distribution, Eq.~(\ref{ellipticpower2d}), encodes the
information about the eccentricity distribution into two parameters
which are plotted in Fig.~\ref{parameters} 
as a function of centrality for the IP Glasma and Monte Carlo Glauber
models. 
As explained above, the two parameters cannot be disentangled 
for very central collisions --- in practice, the fitting procedure returns a very 
large error on each parameter: therefore we 
exclude the most central ($0-5\%$) bin. 
Panel (a) also displays the values of $\alpha$ obtained by fitting
the distribution of the {\it triangularity\/}  $\varepsilon_3$ with
the Power distribution Eq.~(\ref{power}).
The power parameter $\alpha$ increases towards central collisions.  This is expected, since $\alpha$ is typically
proportional to the system size. 
In the Monte Carlo Glauber model, $\alpha$ is approximately
proportional to the number of participant nucleons $N_{\rm part}$. 

The ellipticity $\varepsilon_0$, on the other hand, smoothly increases
with centrality percentile, and is somewhat larger for the IP-Glasma
than for the Glauber model, in line with the expectations that
saturation-inspired models predict a larger eccentricity than
Glauber models~\cite{Lappi:2006xc}. 
For the Monte Carlo Glauber model, we also show on the same plot 
the reaction plane eccentricity $\varepsilon_{\rm{RP}}$: we can either
calculate it directly in the Monte Carlo Glauber model (full line) or
estimate it using Eq.~(\ref{epsRP}) below derived from the Elliptic Power
distribution (dotted line). 
It is close to the Glauber $\varepsilon_0$ up to mid-centrality. The difference between 
$\varepsilon_0$ and $\varepsilon_{\rm{RP}}$ is nevertheless much larger
than predicted by the Elliptic Power distribution. 
This can be attributed to the fact that the Elliptic Power 
distribution does not reproduce all the fine structure of the
two-dimensional distribution (Fig.~\ref{2d} (a)), even though it
provides a very good fit to the distribution of $\varepsilon_2$ 
(Figs.~\ref{1d} (a) and \ref{histo}).

\subsection{Fluctuations}

\begin{figure}
\includegraphics[width=\linewidth]{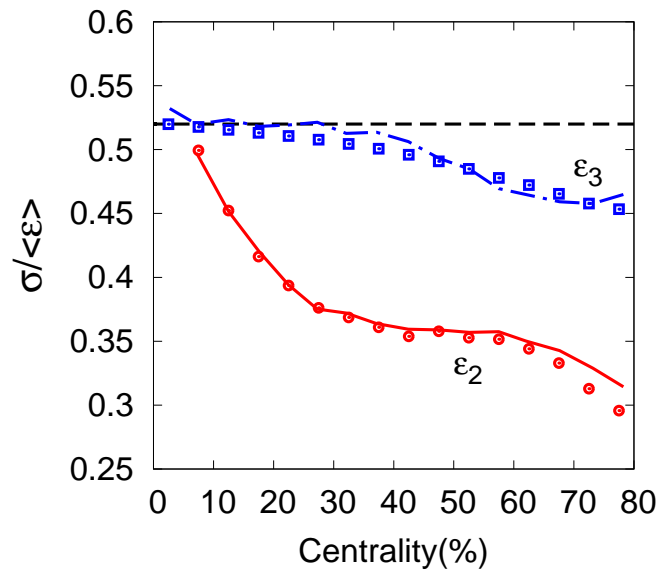}
\caption{ (Color online) 
Relative fluctuations, as defined by
Eq.~(\ref{sigmaovermean}), as a function of collision centrality. 
Lines: Monte Carlo Glauber results for $\varepsilon_2$  (full)
and $\varepsilon_3$ (dash-dotted). 
Symbols are results using the Elliptic Power distribution  Eq.~(\ref{ellipticpower}) for $\varepsilon_2$ (circles), 
and the Power distribution Eq.~(\ref{power}) for $\varepsilon_3$ (squares), with parameters fitted to the 
histograms. 
The horizontal dashed line indicates the value $\sqrt{4/\pi-1}$ corresponding to 
pure Gaussian fluctuations (Eq.~(\ref{BG}) with $\varepsilon_0$).} 
\label{sigovermean}
\end{figure}

A standard measure of eccentricity fluctuations is the ratio of the
standard deviation to the mean~\cite{Miller:2007ri,Sorensen:2006nw}:
\begin{equation}
\label{sigmaovermean}
\frac{\sigma_{\varepsilon_n}}{\langle\varepsilon_n\rangle}=
\frac{\sqrt{\langle\varepsilon_n^2\rangle-\langle\varepsilon_n\rangle^2}}
{\langle\varepsilon_n\rangle},
\end{equation}  
where angular brackets denote an average over events in a centrality class. 
We now check that the Elliptic Power
distribution, fitted to the histogram of $\varepsilon_2$, 
correctly reproduces the magnitude of eccentricity fluctuations.

The ratio Eq.~(\ref{sigmaovermean})  is presented in Fig.~\ref{sigovermean} for $\varepsilon_2$ and 
$\varepsilon_3$. For central collisions,  it approaches 
$\sqrt{4/\pi-1}\simeq 0.52$~\cite{Broniowski:2007ft} for both $\varepsilon_2$
and $\varepsilon_3$, which is the value given by Eq.~(\ref{BG}) for $\varepsilon_0=0$. 
For more peripheral collisions, relative eccentricity fluctuations decrease very mildly
for $\varepsilon_3$, and more strongly for $\varepsilon_2$. 
For $\varepsilon_3$, this mild decrease is captured by fitting with the Power distribution, 
Eq.~(\ref{power}). 
The mean of the Power distribution is given by 
\begin{equation}
\label{meanpower}
\langle\varepsilon_3\rangle=
\frac{\sqrt{\pi}\,\Gamma(\alpha+1)}{2\,\Gamma(\alpha+\frac{3}{2})},
\end{equation}
while the mean square is~\cite{Yan:2013laa}  $\langle\varepsilon_3^2\rangle=1/(\alpha+1)$. 

The mean of the Elliptic Power distribution, Eq.~(\ref{ellipticpower}), 
must be calculated numerically as a function of $\varepsilon_0$ and $\alpha$. 
The eccentricity fluctuations from the Elliptic Power distribution closely match the Monte Carlo 
Glauber result in Fig.~\ref{sigovermean}. 

\subsection{Cumulants}

\begin{figure}
\includegraphics[width=\linewidth]{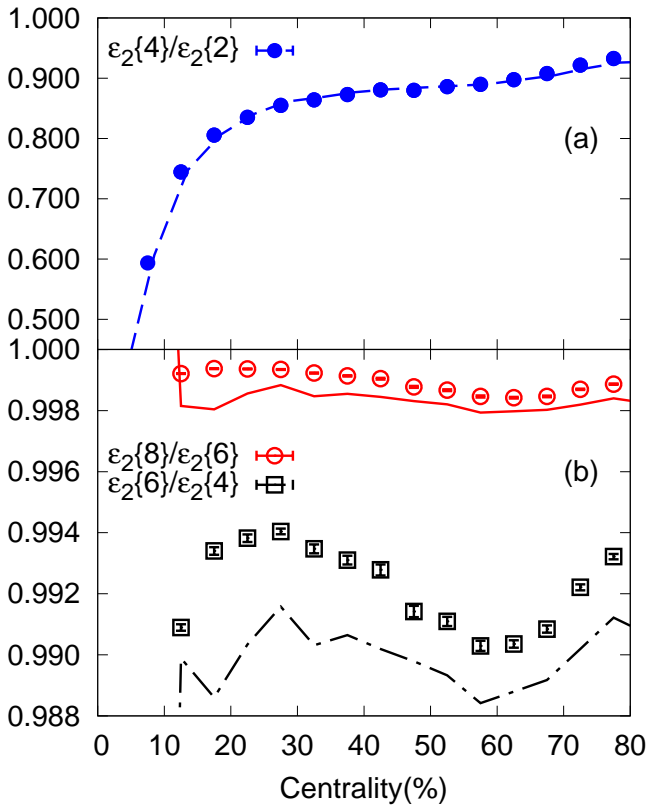}
\caption{ (Color online) 
(a): 
Cumulant ratio $\varepsilon_2\{4\}/\varepsilon_2\{2\}$.
(b): 
Cumulant ratios
$\varepsilon_2\{6\}/\varepsilon_2\{4\}$ and 
$\varepsilon_2\{8\}/\varepsilon_2\{6\}$.
Lines are Monte Carlo Glauber results. 
Symbols are 
results using the Elliptic Power distribution Eq.~(\ref{cumulantsEP}) 
with parameters fitted to the distribution of $\varepsilon_2$. 
}
\label{fig:cumulants}
\end{figure}

More detailed information about eccentricity fluctuations is contained in moments
or cumulants of the distribution. 
The moment of order $k$ is defined as $\langle(\varepsilon_n)^k\rangle$. 
Often, one solely uses {\it even\/} moments of the distribution 
$\langle(\varepsilon_n)^{2k}\rangle$, because the corresponding moments of the
distribution of anisotropic flow are directly accessible through
cumulant analyses~\cite{Borghini:2001vi,Bilandzic:2010jr}.
The first eccentricity 
cumulants~\cite{Miller:2003kd,Bzdak:2013rya} are defined by:
\begin{eqnarray}
\label{cumulants}
\varepsilon_n\{2\}&\equiv&\langle\varepsilon_n^2\rangle^{1/2}\cr
\varepsilon_n\{4\}&\equiv&\left(2\langle\varepsilon_n^2\rangle^2-\langle\varepsilon_n^4\rangle\right)^{1/4}\cr
\varepsilon_n\{6\}&\equiv&\left(\frac{\langle\varepsilon_n^6\rangle-9\langle\varepsilon_n^2\rangle\langle\varepsilon_n^4\rangle
+12\langle\varepsilon_n^2\rangle^3}{4}\right)^{1/6}.
\end{eqnarray}

Figure~\ref{fig:cumulants} displays ratios of successive cumulants obtained 
in the Monte Carlo Glauber calculation and using the Elliptic Power distribution, Eq.~(\ref{cumulantsEP}). 
$\varepsilon_2\{4\}/\varepsilon_2\{2\}$ increases from central to peripheral collisions. 
It is smaller than unity by definition. 
Higher order ratios  $\varepsilon_2\{6\}/\varepsilon_2\{4\}$ 
and $\varepsilon_2\{8\}/\varepsilon_2\{6\}$
are exactly equal to 1 for the Bessel-Gaussian distribution. 
The Monte Carlo Glauber calculation gives ratios slightly smaller than unity, with 
a nontrivial centrality dependence. 
These nontrivial features are reproduced by the Elliptic Power distribution.

\section{Conclusions}

We have introduced a new parametrization of the eccentricity
distribution in nucleus-nucleus collisions. Like the previously used
Bessel-Gaussian parametrization, it is a two-parameter distribution,
but it describes peripheral collisions much better. 
This is due to the correct implementation of the constraint that the
eccentricity must be smaller than unity in all events. 
The consequence of our result is that any model of initial-state
eccentricities can be characterized by two numbers for each
centrality: the ellipticity $\varepsilon_0$, which corresponds
closely to the reaction-plane eccentricity; 
the power parameter $\alpha$, which governs the magnitude of
fluctuations and scales like the number of participants in the Glauber model. 

Since elliptic flow is essentially proportional to the initial
eccentricity~\cite{Niemi:2012aj}, our result can be applied~\cite{Yan2014} 
to the distribution of elliptic flow values, which has been measured
recently at the LHC~\cite{Aad:2013xma,Jia:2013tja}. 
The Elliptic Power distribution could also be used as a kernel in the unfolding
procedure which is used to eliminate finite multiplicity fluctuations~\cite{Alver:2007qw}. 
It could also be used in fitting the distribution of the flow 
vector~\cite{Barrette:1994xr,Voloshin:1994mz,Adler:2002pu}.
We expect it to give a better result than the Bessel-Gaussian distribution, 
which has been found to be not precise for peripheral collisions~\cite{Aad:2013xma}.

\begin{acknowledgments}
We thank Hiroshi Masui for carrying out the Monte Carlo Glauber
calculations which inspired this work, Bj\"orn Schenke for the IP
Glasma results, C. Loizides for the new version of the PHOBOS
Glauber, M. Luzum and S. Voloshin for extensive discussions and suggestions. 
In particular, we thank S. Voloshin for suggesting the name Elliptic Power, and for useful comments on the manuscript. 
JYO thanks the MIT LNS for hospitality. 
LY is funded  by the European Research Council under the 
Advanced Investigator Grant ERC-AD-267258. AMP was supported by the Director, Office of Science, of the U.S. Department of Energy.
\end{acknowledgments}

\appendix
\section{Mathematical properties of the Elliptic Power distribution}
\label{sec:appendix}

The two-dimensional Elliptic Power distribution Eq.~(\ref{ellipticpower2d})
is normalized to unity on the unit disk if $\alpha>0$ and $-1<\varepsilon_0<1$. 
We choose the convention $\varepsilon_0\ge 0$ throughout this paper. 

For $\alpha\ge 1$, Eq.~(\ref{ellipticpower2d}) has a maximum on the 
$x$-axis for $\varepsilon_x=\varepsilon_{\rm max}$, where
\begin{equation}
\label{max}
\varepsilon_{\rm max}\equiv\frac{\varepsilon_0(1+2\alpha)}{\alpha-1+
\sqrt{(\alpha-1)^2+3\varepsilon_0^2(1+2\alpha)}}.
\end{equation}
But for $0<\alpha<1$, the distribution diverges on the unit circle
$\varepsilon_x^2+\varepsilon_y^2=1$. 

The mean eccentricity in the reaction plane is obtained by integrating
Eq.~(\ref{ellipticpower2d}):
\begin{eqnarray}
\varepsilon_{\rm{RP}}&=&
\int_{-1}^{+1}\varepsilon_xd\varepsilon_x
\int_{-\sqrt{1-\varepsilon_x^2}}^{\sqrt{1-\varepsilon_x^2}}d\varepsilon_y 
p(\varepsilon_x,\varepsilon_y)\cr
&=&
\frac{\alpha+\frac{1}{2}}{\alpha+1}\varepsilon_0(1-\varepsilon_0^2)^{\alpha+\frac{1}{2}}\times \cr
&&
{_2}F_1\left(\alpha+1,\alpha+\frac{3}{2};\alpha+2;\varepsilon_0^2\right),
\label{epsRP}
\end{eqnarray} 
where $_2F_1$ denotes the hypergeometric function. 
In the limit $\varepsilon_0\ll 1$, it simplifies to 
\begin{equation}
\varepsilon_{\rm{RP}}=
\frac{\alpha+\frac{1}{2}}{\alpha+1}\varepsilon_0+{\cal O}(\varepsilon_0^3). 
\label{epsRPsmall}
\end{equation} 
Thus $\varepsilon_{0}$ is slightly bigger than $\varepsilon_{\rm{RP}}$ due to fluctuations. 

We now explain how to evaluate the moments of the Elliptic Power
distribution. Multiplying Eq.~(\ref{ellipticpower2d}) by 
$(1-\varepsilon_x^2+\varepsilon_y^2)^k$ and integrating successively over
$\varepsilon_y$ and $\varepsilon_x$, one obtains
\begin{eqnarray}
\langle (1-\varepsilon_n^2)^k\rangle
&=&\frac{\alpha}{\alpha+k}
\left(1-\varepsilon_0^2\right)^k\times\cr
&&
{_2}F_1\left(k+\frac{1}{2},k;\alpha+k+1;\varepsilon_0^2\right). 
\end{eqnarray} 
Using this equation, one can express analytically the even moments 
$\langle\varepsilon_n^{2k}\rangle$ and the cumulants~\cite{Bzdak:2013rya}
$\varepsilon_n\{2k\}$ as a function of $\alpha$ and $\varepsilon_0$. 
Using the shorthand notation $f_k\equiv \langle
(1-\varepsilon_n^2)^k\rangle$ and inserting into Eq.~(\ref{cumulants}), 
one obtains
\begin{eqnarray}
\label{cumulantsEP}
\varepsilon_n\{2\}&=&(1-f_1)^{1/2} \cr
\varepsilon_n\{4\}&=&(1 - 2 f_1 + 2 f_1^2  - f_2)^{1/4} \cr
\varepsilon_n\{6\}&=&\left(1+\frac{9}{2}
f_1^2-3f_1^3+3f_1(\frac{3}{4}f_2-1)-\frac{3}{2} f_2\right.\cr 
&&\left. -\frac{1}{4}f_3\right)^{1/6}.
\end{eqnarray}

\section{Limiting distribution for fixed $\varepsilon_0>0$ and $\alpha\gg 1$}
\label{sec:asymptotic}

We now study the Elliptic Power distribution Eq.~(\ref{ellipticpower2d})
in the limit  $\alpha\gg 1$, corresponding to a large system. 
To leading order, the distribution is a Gaussian centered around the
intrinsic ellipticity $\varepsilon_0$, see 
Eq.~(\ref{BG2d}). 
We therefore write
$\varepsilon_x=\varepsilon_0+\delta_x$ and treat $\delta_x$ and
$\varepsilon_y$ as small parameters of order $\sigma_x\sim\sigma_y\sim
\alpha^{-1/2}$. Expanding the logarithm of
Eq.~(\ref{ellipticpower2d}) in powers of $\alpha^{-1/2}$ and
exponentiating, one obtains 
\begin{equation}
\label{expansion}
p(\varepsilon_0+\delta_x,\varepsilon_y)=
p_0(\varepsilon_0+\delta_x,\varepsilon_y)\left(1+W_1+W_3\right),
\end{equation} 
where $p_0$ is the Gaussian distribution in Eq.~(\ref{BG2d}), and $W_1$ and $W_3$ are 
perturbations of order $\alpha^{-1/2}$:
\begin{eqnarray}
\label{perturbation}
W_1&\equiv&
\frac{3\varepsilon_0\delta_x}{1-\varepsilon_0^2}\cr
W_3&\equiv&
-\left(\frac{\delta_x^2}{\sigma_x^2}+\frac{\varepsilon_y^2}{\sigma_y^2}\right)
\frac{\varepsilon_0\delta_x}{1-\varepsilon_0^2}.
\end{eqnarray}
$W_1$ is linear, while $W_3$ is cubic in $\delta_x$ and $\varepsilon_y$. 
The linear term $W_1$ shifts the maximum of the distribution,
which is found by
setting $\varepsilon_y\equiv 0$ in Eq.~(\ref{expansion}) and
differentiating with respect to $\delta_x$: 
\begin{equation}
\label{maxnlo}
\varepsilon_{\rm max}=
\varepsilon_0+\frac{3\varepsilon_0(1-\varepsilon_0^2)}{2\alpha}+{\cal O}\left(\frac{1}{\alpha^2}\right).
\end{equation}
Alternatively, this result can be recovered by expanding Eq.~(\ref{max}). 

The cubic term $W_3$  skews the Gaussian and is
responsible for the skewness seen in Fig.~\ref{2d} (b) and Fig.~\ref{1d} (a). 
The linear term $W_1$ can be absorbed by shifting the maximum of the Gaussian 
according to Eq.~(\ref{maxnlo}), therefore the difference
between $\varepsilon_{\rm{RP}}$ and $\varepsilon_{\rm max}$ is solely due
to $W_3$:
\begin{eqnarray}
\label{avnlo}
\varepsilon_{\rm{RP}}&=&\varepsilon_{\rm max}+
\int p_0(\varepsilon_0+\delta_x,\varepsilon_y)W_3 \delta_x d\delta_x d\varepsilon_y\cr
&=&\varepsilon_0-\frac{\varepsilon_0(1-\varepsilon_0^2)}{2\alpha}
+{\cal O}\left(\frac{1}{\alpha^2}\right).
\end{eqnarray}
Comparing with Eq.~(\ref{maxnlo}), one sees that
$\varepsilon_{\rm{RP}}<\varepsilon_{\rm max}$, which is a consequence of
the skewness.
Alternatively, Eq.~(\ref{avnlo}) can be obtained by expanding
Eq.~(\ref{epsRP}). 

The first moments can also be evaluated to first order in
$1/\alpha$. The mean square eccentricity is 
\begin{equation}
\label{mom2}
\varepsilon_n\{2\}^2=\langle\varepsilon_n^2\rangle=
\varepsilon_0^2+\frac{(1-\varepsilon_0^2)\left(1-\frac{3}{2}\varepsilon_0^2\right)}{\alpha}+{\cal O}\left(\frac{1}{\alpha^2}\right).
\end{equation}
When $\varepsilon_0\to 0$, one recovers the result obtained with the
Power distribution~\cite{Yan:2013laa} in the limit $\alpha\gg 1$. When
$\varepsilon_0>\sqrt{\frac{2}{3}}\simeq 0.816$, the correction is
negative, so that rms anisotropy is {\it smaller\/} than $\varepsilon_0$. 
The fourth moment is given by 
\begin{equation}
\label{mom4}
\langle\varepsilon_n^4\rangle=
\varepsilon_0^4+\frac{\varepsilon_0^2(1-\varepsilon_0^2)(4-5\varepsilon_0^2)}{\alpha}+{\cal O}\left(\frac{1}{\alpha^2}\right).
\end{equation}
From Eqs.~(\ref{mom2}) and (\ref{mom4}), one obtains the cumulant
$\varepsilon_n\{4\}$ (see Eq.~(\ref{cumulants})):
\begin{equation}
\varepsilon_n\{4\}=
\varepsilon_0-\frac{\varepsilon_0(1-\varepsilon_0^2)}{4\alpha}+{\cal O}\left(\frac{1}{\alpha^2}\right).
\end{equation}
Note that $\varepsilon_n\{4\}<\varepsilon_0$ for all positive
$\varepsilon_0$ in the limit $\alpha\gg 1$. 
In the limiting case $\varepsilon_0=0$, 
$\varepsilon_n\{4\}^4$ is positive and of order
$\alpha^{-3}$~\cite{Yan:2013laa}. 
Higher order cumulants are all equal to $\varepsilon_n\{4\}$ to order
$1/\alpha$. 

\section{Limiting distribution for fixed $\alpha\varepsilon_0^2$ and $\alpha\gg 1$}
\label{sec:asymptotic2}

We now consider a different asymptotic expansion introduced in
\cite{Alver:2008zza}, where one treats $\alpha$ as a large parameter
and $\varepsilon_0$ as a small parameter, keeping the product
$\alpha\varepsilon_0^2$ fixed. 
The only difference with the asymptotic expansion carried out in 
Appendix~\ref{sec:asymptotic} is that we also treat 
$\varepsilon_0$ as a small parameter of order $\alpha^{-1/2}$. 
Therefore the perturbations $W_1$ and $W_3$ in
Eq.~(\ref{perturbation}) are of order $\alpha^{-1}$. 
For sake of consistency, one must carry out the whole expansion to
that order. One obtains
\begin{equation}
\label{expansion2}
p(\varepsilon_0+\delta_x,\varepsilon_y)=
p_0(\varepsilon_0+\delta_x,\varepsilon_y)\left(1+W_1+W_3+W_4\right),
\end{equation} 
where $W_1$ and $W_3$ are defined in Eq.~(\ref{perturbation}) 
and $W_4$ is a new quartic term: 
\begin{eqnarray}
\label{perturbation2}
W_1&\equiv&
3\varepsilon_0\delta_x\cr
W_3&\equiv&
-2\alpha\varepsilon_0\delta_x\vec\delta^2\cr
W_4&\equiv&
-\frac{\alpha}{2}(\vec\delta^2)^2+\vec\delta^2,
\end{eqnarray}
where we have introduced the shorthand notation
$\vec\delta^2\equiv\delta_x^2+\varepsilon_y^2$ and simplified the
expressions of $W_1$ and $W_3$ using 
Eq.~(\ref{sigmaxy}) and $\varepsilon_0\ll 1$. 
In the isotropic case $\varepsilon_0=0$, both $W_1$ and $W_3$ vanish
and only $W_4$ contributes. 

The mean square eccentricity is given by 
\begin{equation}
\varepsilon_n\{2\}^2=\varepsilon_0^2+\frac{1}{\alpha}-\frac{5\varepsilon_0^2}{2\alpha}-\frac{1}{\alpha^2}+{\cal
  O}\left(\frac{1}{\alpha^3}\right), 
\end{equation}
where the first two terms are the leading order terms, of order
$1/\alpha$, and the two next terms are corrections of order
$1/\alpha^2$. The first three terms are present in 
Eq.~(\ref{mom2}), while the last term is the contribution of the
quartic perturbation $W_4$ in Eq.~(\ref{perturbation2}). 
Similarly, one can expand the cumulant $\varepsilon_n\{4\}$, to give for the fourth power:
\begin{equation}
\varepsilon_n\{4\}^4=\varepsilon_0^4-\frac{\varepsilon_0^4}{\alpha}+\frac{8\varepsilon_0^2}{\alpha^2}+\frac{2}{\alpha^3}
+{\cal
  O}\left(\frac{1}{\alpha^4}\right), 
\end{equation}
where the first term is the leading term, of order $1/\alpha^2$,  and
the next three terms are corrections of order $1/\alpha^3$. 
In the isotropic case $\varepsilon_0=0$, the exact result is
$\varepsilon_n\{4\}^4=2/[(\alpha+1)^2(\alpha+2)]$~\cite{Yan:2013laa},
which reduces to
$\varepsilon_n\{4\}^4\simeq 2/\alpha^3$ for $\alpha\gg 1$, in agreement
with the above result.

Ratios of cumulants are given to leading order by: 
\begin{eqnarray}
\frac{\varepsilon\{4\}}{\varepsilon\{2\}}&=&\sqrt{\frac{\alpha\varepsilon_0^2}{1+\alpha\varepsilon_0^2}}+{\cal
  O}\left(\frac{1}{\alpha}\right)\cr
\frac{\varepsilon\{6\}}{\varepsilon\{4\}}&=&1-\frac{1+\alpha\varepsilon_0^2}{2(\alpha\varepsilon_0^2)^2\alpha}+{\cal
  O}\left(\frac{1}{\alpha^2}\right)\cr
\frac{\varepsilon\{8\}}{\varepsilon\{6\}}&=&1-\frac{1}{22(\alpha\varepsilon_0^2)\alpha}+{\cal
  O}\left(\frac{1}{\alpha^2}\right).\label{phobos}
\end{eqnarray}

\end{document}